\documentclass[english,russian,12pt,twoside]{article}
\usepackage{amssymb,amsmath}
\usepackage[russian]{babel}
\usepackage[cp866]{inputenc}
\usepackage{graphicx}
\usepackage{amscd}
\usepackage{epsfig}
\begin{document}
УДК 539.3

MSQ 74B05,35E05

\begin{center}
{\bf ОБОБЩЕННЫЕ РЕШЕНИЯ УРАВНЕНИЙ ЛАМЕ \\ В СЛУЧАЕ БЕГУЩИХ НАГРУЗОК. УДАРНЫЕ ВОЛНЫ}
\end{center}
\vspace{3mm}
 \centerline{\textbf{ Л.А. Алексеева} }
\vspace{3mm} \centerline{\textit{Институт Математики ИМИМ МОН РК}}
 \centerline{\textit {050010, Алматы, ул. Пушкина,125, Казахстан}}
 \centerline{ alexeeva@math.kz}
 \vspace{5mm}

Математическое моделирование разнообразных
процессов, связанных с передвижением транспорта в различных средах, либо перемещением
транспортируемых грузов в тоннелях и трубопроводах различного назначения приводит
 к решению краевых задач механики сплошных сред  в классе
"бегущих" функций, параметрических и автомодельных по ряду переменных. Параметр задачи - скорость
движения источника возмущений в среде - существенно влияет на тип уравнений движения, который
зависит от скоростей распространения волн в средах, так называемых \emph{звуковых скоростей}. Их
может быть несколько в зависимости от вида волн. Тип дифференциальных уравнений, описывающих
движение среды, меняется в зависимости от отношения скорости источника возмущений к звуковым
скоростям (чисел Маха). Поэтому  приходится строить решения систем уравнений эллиптического,
гиперболического или смешанного типов. При этом, как известно, гладкость решений существенно зависит
 от гладкости граничных функций, которые зависят от типа действующих поверхностных нагрузок и массовых сил.

 Для
 физических задач типичными являются ударные воздействия, сосредоточенные на поверхностях и в
 точках силы и т.п., которые не описываются  гладкими функциями.
Удобный метод для решения таких задач дает аппарат теории обобщенных функций [1,2], который позволяет
существенно расширить класс изучаемых процессов, используя сингулярные обобщенные функции для моделирования
наблюдаемых явлений, в особенности описываемых гиперболическими или смешанными уравнениями,
для которых математическая теория краевых задач пока еще недостаточно развита.
Основные идеи  метода обобщенных  функций для решения краевых задач для волновых уравнений в N-мерных
пространствах в классе бегущих решений рассмотрены в [3,4].  Здесь этот метод
используется для  построения обобщенных решений уравнения Ламе, описывающих движение упругой среды,
при дозвуковых, транс-  и сверхзвуковых скоростях движения источника возмущений. Рассмотрены ударные волны,
которые возникают в среде при сверхзвуковых источниках возмущений, и предложен метод определения
условий на скачки решений и их производных на фронтах ударных волн.
 \vspace {3mm}

\textbf{1.  Уравнения движения упругой среды. Ударные волны}. Рассмотрим изотропную упругую среду,
 заданную параметрами Ламе $\lambda ,\mu $,
плотностью $\rho$. Обозначим $ u_i $ - компоненты вектора перемещений $u$; $\sigma _{ij}
,\varepsilon _{ij} $ - компоненты тензора напряжений и деформаций, связанные законом Гука [5]:
\begin{equation}\label{(1)}
 \sigma _{ij}  = \lambda \,div\,u\,\delta _{ij}  + 2\mu \,{\kern 1pt}
\varepsilon _{ij},
\end{equation}
\begin{equation}\label{(2)}
\varepsilon _{ij}  = \frac{1}{2}\left\{ {\frac{{\partial u_i }}{{\partial x_j }} + \frac{{\partial
u_j }}{{\partial x_i }}} \right\},\quad\quad i,j,k = \overline {1,N} .
\end{equation}
Здесь $(x_1 , \ldots x_N )$ -- лагранжевы декартовы координаты точек упругой среды.
При $N=2$ деформация плоская:  $u_j  = u_j (x_1 ,x_2 ,t),\,\,j = 1,2,u_3  = 0;$ при $N=3$ -
пространственная:  $u_j  = u_j (x_1 ,x_2 ,x_3,t),\,\,j = 1,2,3 $.

Уравнения движения сплошной среды
    \begin{equation}\label{(3)}
\frac{{\partial \sigma _{ij} }}{{\partial x_j }} + \rho G_i  = \rho \frac{{\partial ^2 u_i
}}{{\partial t^2 }}
\end{equation}
для упругой среды, учетом (1), (2), они приводятся к виду:
    \begin{equation}\label{(4)}
L_i^j \left( {\partial _x ,\partial _t } \right)u_j (x,t) + G_i (x,t) = 0,
\end{equation}
где $L_I^J $ - дифференциальный оператор Ламе:
\[
L_i^j \left( {\partial _x ,\partial _t } \right) = (c_1^2  - c_2^2 )\frac{\partial }{{\partial x_i
}}\frac{\partial }{{\partial x_j }} + \delta _i^j \left( {c_2^2 \Delta _N  - \frac{{\partial ^2
}}{{\partial t^2 }}} \right),
\]
$c_1  = \sqrt {\left( {\lambda  + 2\mu } \right)/\rho } , \,c_2  = \sqrt {\mu /\rho } $ -- скорости
распространения объемных и сдвиговых волн в упругой среде ($c_1>c_2$), $G_i $ -- декартовы координаты объемной
силы, $\delta _i^j = \delta _{ij} $  -- символ Кронекера, $\Delta _N $ --  оператор Лапласа в $R^N$.
В (1) и далее по повторяющимся индексам проводится суммирование от 1 до $N$ (подобно тензорной
свертке).

Система уравнений (4) достаточно подробно исследована в работах  Г.И.Петрашеня[6].
 В силу положительной определенности упругого потенциала среды, она является строго
гиперболической. Детерминант ее характеристической матрицы
\[
 L =
\left\{ {L_i^j \left( {\xi ,\omega } \right)} \right\} = \left\{ {\left( {c_1^2  - c_2^2 }
\right)\xi _i \xi _j  + \left( {c_2^2 \left\| \xi  \right\|^2  - \omega ^2 } \right)\delta _i^j }
\right\}
\]
 имеет $2N$ (с учетом кратности) действительных корней:  6 при $N=3 $
 $\left( \omega /\left\| \xi  \right\| =  \pm c_1 , \pm c_2 , \pm c_2  \right)
$, и 4 при $N=2$:  $\left( \omega /\left\| \xi  \right\| =  \pm c_1 , \pm c_2  \right)
$. Здесь  $\xi  = (\xi _1 , \ldots ,\xi _N ),\,\left\| \xi  \right\| = \sqrt {\sum\limits_{j = 1}^N
{\xi _j^2 } } $.

Гиперболические системы допускают разрывные по производным решения. Поверхность разрыва  $F$  в
$R^{N + 1}  = R^N  \times t$,  $\left( { - \infty  < t < \infty } \right)$, совпадает с
характеристической поверхностью системы.  Ей соответствует волновой фронт $F_t $, который движется
в пространстве  $R^N $ с течением времени.

 Пусть $\nu  = \left\{
{\nu _1 , \ldots ,\nu _N ,\nu _t } \right\}$ -- единичный вектор нормали к $F$ в $R^{N + 1} $,
удовлетворяющий характеристическому уравнению:
\begin{equation}\label{(5)}
\det \left\{ {\left( {c_1^2  - c_2^2 } \right)\nu _i \nu _j  + \delta _{ij} \left( {c_2^2 \left\|
\nu  \right\|_N^2  - \nu _t^2 } \right)} \right\} = 0,\quad\left\| \nu  \right\|_N  = \sqrt
{\sum\limits_{k = 1}^N {\nu _k^2 } ,}
\end{equation}
В силу гиперболичности системы (4),  это уравнение имеет корни:
          \begin{equation}\label{(6)}
\nu _t  =  \pm c_j \left\| \nu  \right\|_N ,\quad j = 1,2.
\end{equation}
Поверхность $F_t $ движется в $R^N $  со скоростью $V$, которая, как известно, равна
              \begin{equation}\label{(7)}
V =  - \nu _t /\left\| \nu  \right\|_N .
\end{equation}
 Из (6),(7) следует, что $F_t $  движется в пространстве
$R^N $   с одной из звуковых скоростей: $V = c_1 $ или $ V =c_2$.

 Введем \emph{волновой} вектор  $m = (m_1 ,
\ldots ,m_N )$ --  это единичный вектор нормали к фронту  $F_t $  в $R^N $   при фиксированном $t$,
направленный в сторону ее распространения. В силу (7)
      \begin{equation}\label{(8)}
m_j  = \frac{{\nu_j }}{{\left\| \nu  \right\|_N }} =  - V\nu_j /\nu_t.
\end{equation}
Обозначим $\nu_t  = \nu_{N + 1} $. Требование непрерывности перемещений при переходе через волновой
фронт, связанное с сохранением сплошности среды,-
    \begin{equation}\label{(9)}
\left[ u \right]_{F_t }  = 0,
\end{equation}
 приводит к известным кинематическим условиям совместности решений на волновых фронтах [6]:
    \begin{equation}\label{(10)}
\left[ {m_j \frac{{\partial u_i }}{{\partial t}} + V\frac{{\partial u_i }}{{\partial x_j }}}
\right]_{F_t }  = 0,\quad i,j = \overline {1,N} ,
\end{equation}
(условие непрерывности касательных производных  на $F_t $ ). Помимо этого из (4) следуют
динамические условия совместности решений на фронтах,  эквивалентные закону сохранения импульса в
его окрестности [6]:
\begin{equation}\label{(11)}
\left[ {\sigma _{ij} m_j  + \rho V\frac{{\partial u_i }}{{\partial t}}} \right]_{F_t }  = 0.
\end{equation}

О п р е д е л е н и е. Волну назовем \emph{ударной}, если скачок напряжений на фронте волны
конечен: $e_i \left[ {\sigma _{ij} m_j } \right]_{F_t }  \ne 0$, где  $e_i $ -- орты координатных
осей.   Если   $\left[ {\sigma _{ij} m_j } \right]_{F_t } = 0 $,   то волна \emph{слабая
ударная}. Если $\left[ {\sigma _{ij} m_j } \right]_{F_t }  = \infty $, волна \emph{сильная
ударная}.

На  фронтах ударных волн происходит скачок скоростей. На фронтах слабых
 ударных волн нет скачка скоростей, но вторые производные решений разрывны.
Случай сильных ударных волн (в данном определении) в реальных средах не реализуется, т.к. при
больших скачках напряжений среда разрушается и перестает быть упругой. Однако сильные ударные волны
в упругих средах играют важную теоретическую роль при построении решений различных краевых задач. К
таковым, в частности, относятся фундаментальные решения уравнений (4). \vspace {3mm}

\textbf{2.  Бегущие решения уравнений Ламе. Числа Маха}. Пусть  сила, действующая в среде,
движется с постоянной скоростью $c$ вдоль координатной оси $X_N $ ( для удобства выкладок,
противоположно ее направлению) и в подвижной системе координат не зависит от $t$:
            \begin{equation}\label{(12)}
G  = G (x_1 , \ldots ,x_N  + ct)
\end{equation}
Будем искать решения (4) такой же структуры:
            \begin{equation}\label{(13)}
u  = u(x_1 , \ldots ,x_N  + ct)
\end{equation}
которые назовем \emph{бегущими.}

Введем подвижную систему координат $x' = (x'_1 , \ldots ,x'_N )=(x_1 , \ldots ,x_N  + ct)$. В новых
переменных уравнения движения  имеют вид:
 \begin{equation}\label{(14)}
\left\{ {\left( {c_1^2  - c_2^2 } \right)\frac{{\partial ^2 }}{{\partial x'_i \partial 'x_j }} +
\left( {c_2^2 \Delta  - c^2 \frac{{\partial ^2 }}{{\partial x_N'^2 }}} \right)\delta _j^i }
\right\}u_i  + G_j  = 0
\end{equation}
В силу гиперболичности исходной системы, уравнения (14) также могут иметь разрывные решения. Пусть
$F$ поверхность разрыва в пространстве переменных $x'$, где она неподвижна, и  движущаяся с одной
из звуковых скоростей $V = c_1 ,c_2 $ в пространстве переменных $(x_1 , \ldots ,x_N )$. Из (7)
следует, что $V = ch_N $,  где $h=(h_1 , \ldots ,h_N )$ - единичная нормаль к $F$ в $R^N $. Значит,
поскольку $c = c_j /h_N $  и $\left| {h_N } \right| \le 1$,  такие поверхности могут возникнуть
лишь при сверхзвуковых скоростях: $c \ge c_j $.

Назовем скорость $c$ \emph{дозвуковой}, если $c < c_2 $; \emph{межзвуковой}, если $c_2  < c < c_1 $
и \emph{сверхзвуковой}, если  $c > c_1 $. Скорость называется \emph{первой} или \emph{второй
звуковой}, если $c = c_j , \,\, j = 1,2$ соответственно.

Перепишем уравнение (14), поделив его на $c^2$:
\begin{equation}\label{(15)}
A_j^i \left( {\frac{\partial }{{\partial x}}} \right)u_i  = \left\{ {\left( {M_1^{ - 2}  - M_2^{ -
2} } \right)\frac{{\partial ^2 }}{{\partial x'_i \partial 'x_j }} + \left( {M_2^{ - 2} \Delta -
\frac{{\partial ^2 }}{{\partial x_N'^2 }}} \right)\delta _j^i } \right\}u_i  = -g_j
\end{equation}
Здесь $g_j=c^{-2}G_j$,  $M_j  = c/c_j $ -- числа Маха $(M_1<M_2)$. При $M_j  < 1\,(j = 1,2)$ нагрузка дозвуковая,
система уравнений  эллиптического типа;  если нагрузка сверхзвуковая,
$M_j  > 1\,\,(j = 1,2)$  система становится гиперболической;
если скорость межзвуковая (\emph{трансзвуковая}), $M_1  < 1,\,M_2  > 1$ и тип уравнений гиперболо-эллиптический.
При звуковых скоростях уравнения параболо-эллиптические, если $M_2  = 1$ , а при  $M_1  = 1$
становятся параболо-гиперболическими. Покажем это далее при построении фундаментальных решений уравнений (4).

Кинематические и динамические условия совместности решений на разрывах в пространстве переменных  ,
как следует из  (9) - (11) с учетом (13), примут вид:
        \begin{equation}\label{(16)}
\left[ u \right]_F  = 0,
\end{equation}
        \begin{equation}\label{(17)}
\left[ {h_z \;u_i ,_j  - h_j \,u_i ,_N } \right]_F  = 0,
\end{equation}
                 \begin{equation}\label{(18 )}
\left[ {\;\sigma _{ij} h_j  - \rho \,c^2 \;h_N \,\,u_i ,_N } \right]_F  = 0.
\end{equation}
Здесь и часто далее дифференцирование по $x_j$ обозначается индексом $j$ после запятой в обозначении функции.

О п р е д е л е н и е. При   $c > c_2 $  будем называть  решение уравнений (14)
\emph{классическим,} если оно непрерывно, дважды дифференцируемо всюду, за  исключением волновых
фронтов, число которых конечно на любом замкнутом подмножестве в $R^N$, на которых удовлетворяются
условия на скачки (17), (18).

Рассмотрим уравнения (14) и его решения на пространстве обобщенных функций. \vspace {3mm}

\textbf{3.  Обобщенные вектор-функции и их производные.}
 Вначале дадим некоторые определения для обобщенных вектор-функций,
  которые распространяют известную теорию обобщенных функций  [1,2] на пространство вектор-функций.

Пространство финитных бесконечно - дифференцируемых  $M$-мерных вектор-функций на $R^N $ обозначим
$ D_M (R^N ) = \left\{ {\varphi (x) = (\varphi _1 (x),...,\varphi _M (x)),\,\;\varphi _k \in D(R^N
),\;k = \overline {1,M} } \right\} $,  а через $ D_M' (R^N ) = \left\{ {\hat f = (\hat f_1
,...,\hat f_M ),\,{\kern 1pt} \hat f_k  \in D'(R^N ),\;k = \overline {1,M} } \right\} $ -
соответствующее ему пространство обобщенных вектор-функций - непрерывных линейных функционалов на
$D_M (R^N )$:
$(\hat f,\varphi ) = \sum\limits_{k = 1}^M {(\hat f_k ,\varphi _k
)} ,\quad \varphi _k (x) \in D(R^N ) .$
 Если   $\hat f$ -- локально
интегрируемая вектор-функция, то ей соответствует регулярная обобщенная вектор-функция, которая
определяет следующий линейный
 функционал:
 \begin{equation}\label{(19)}
(\hat f,\varphi ) = \sum\limits_{k = 1}^M {\int\limits_{R^N } {f_k (x)\varphi _k )} } (x)dx_1
...dx_N ,\quad \varphi _k (x) \in D(R^N ).
 \end{equation}
 Если   $\hat f$  нельзя представить в интегральном виде (20), то  $\hat f$   - сингулярная
вектор-функция.

 В качестве примера приведем сингулярную обобщенную функцию
 $\alpha (x)\delta _S (x)$, которую называют "\emph{простым слоем на поверхности S}" :
$$(\alpha (x)\delta _S (x),\varphi (x)) =
 \sum\limits_{k = 1}^M {\int\limits_{S(x)} {\alpha _k (x)\varphi _k } } (x)dS(x). $$
       Здесь вектор-функция $\alpha=\{\alpha _1 (x),...,\alpha _M (x)\}$ определена и локально интегрируема на поверхности $S \subset R^N $.  Заметим,
что  в отличие от (19), интеграл берется не по  $R^N $, а по поверхности $S$. Далее будем
предполагать, что $DimS=N-1$,  $S$  непрерывна и имеет непрерывную нормаль $n(x) = (n_1 (x),...,n_N
(x)),\;\,\left\| n \right\| = 1.$

Дифференцирование на $D_M' (R^N )$  определяем, используя производную обобщенной функции:
 \begin{equation}\label{(20)}
(\partial ^a \hat f,\varphi ) = ( - 1)^{\left| a \right|}
 (\hat f,\partial ^a \varphi ) = ( - 1)^{\left| a \right|}
 \sum\limits_{k = 1}^M {(\hat f_k ,\partial ^a \varphi _k )}.
  \end{equation}
  где мультииндекс $a = (a_1 ,...,a_N )$ определяет
частную производную $$\partial ^a  = \frac{{\partial ^{\left| a \right|} }}{{\partial x_1^{a_1 }
...\partial x_N^{a_N } }},\;\left| a \right| = \sum\limits_{k = 1}^N {a_k } .$$

Если $f(x)$ дифференцируема всюду, кроме поверхности S, то, как известно [1,2],
\begin{equation}\label{(21)}
\hat f,_j  =  f,_j  + n_j \left[ f \right]_S \delta _S (x)
  \end{equation}
Здесь слева стоит обобщенная производная по $x_j$ в смысле определения (20), а справа первое слагаемое -
обычная производная. Очевидно, что (21) справедливо и для обобщенных вектор-функций. Далее будем
называть обобщенные вектор-функции просто \emph{обобщенными} функциями.

Рассмотрим на $D_M' (R^N )$линейный дифференциальный оператор с постоянными коэффициентами
$$L\left( {\partial _x } \right) = \left\{ {L_{ij} \left( {\partial _x } \right)} \right\}_{M
\times M}  = \left\{ {\sum\limits_a {C_i^{ja} } \partial ^a } \right\}_{M \times M}$$

О п р е д е л е н и е. Назовем $\hat u$ \emph{обобщенным решением} дифференциального уравнения
 \[ L\left( {\partial _x } \right)\hat
u = \left\{ {L_i^j \left( {\partial _x } \right)\hat u_j } \right\}_{M \times 1}  = \left\{ {\hat
f_i } \right\}_{M \times 1} = \hat f,
\]
если
\[
\left( {L\left( {\partial _x } \right)\hat u,\varphi } \right) = \left( {L_{ij} \left( {\partial _x
} \right)\hat u_j ,\varphi _i } \right) = \left( {\hat u_j ,L_{ji}^* \left( {\partial _x }
\right)\varphi _i } \right) = \left( {\hat f,\varphi } \right),\quad \forall \varphi  \in D_M (R^N
).
\]
Здесь сопряженный оператор $L_{ij}^* \left( {\partial _x } \right) = \sum\limits_a {( - 1)^{\left|
a \right|} C_j^{ia} } D^a. $

О п р е д е л е н и е. Матрица $\hat U(x)$ называется матрицей фундаментальных решений, если она
удовлетворяет матричному уравнению
\begin{equation}\label{(22)}
L\left( {\partial _x } \right)\hat U = \left\{ {L_{ij} \left( {\partial _x } \right)\hat U_j^k (x)}
\right\}_{M \times M}  = \delta (x)\left\{ {\delta _i^k } \right\}_{M \times M}
\end{equation}
где $\delta (x)$ сингулярная $\delta $-функция: \[ \left( {\delta (x)\delta _i^k ,\varphi _k (x)}
\right) = \left( {\delta (x),\delta _i^k \varphi _k (x)} \right) = \left( {\delta (x),\varphi _i
(x)} \right) = \varphi _i (0).
\]

    Ясно, что фундаментальные решения определяются с точностью до
решения однородного уравнения.

    Т е о р е м а 1. \emph{Если существует свертка
\begin{equation}\label{(23)}
\hat u = \hat U*\hat f = \left\{ {\hat U_j^k (x)*\hat f_k (x)} \right\}_{M \times 1},
\end{equation}
то $\hat u$ является обобщенным решением уравнения}
                        $L\left( {\partial _x } \right)\hat u = \hat f.$

Д о к а з а т е л ь с т в о. В силу свойства дифференцирования свертки и свертки с $\delta (x)$ [1,2
]:
\[
 L\left( {\partial _x }
\right)\hat u = L\left( {\partial _x } \right)\left( {\hat U*\hat f} \right) = \left( {L\left(
{\partial _x } \right)\hat U} \right)*\hat f = \left\{ {\delta _i^k \delta (x)*\hat f_k }
\right\}_{M \times 1}  = \hat f.
\]
Ч.т.д.

Если входящие в свертку обобщенные функции являются регулярными, то они записываются в интегральной
форме вида:
\begin{equation}\label{(24)}
\hat u = \hat U*\hat f = \left\{ {\int\limits_{R^N } {U_j^k (x - y)} f_k (y)dy_1 ...dy_N }
\right\}_{M \times 1}
\end{equation}

Итак, зная матрицу фундаментальных решений системы дифференциальных уравнений, можно строить их
обобщенные решения для разных правых частей из класса обобщенных функций. Как легко видеть в (24),
решение представляется в виде суперпозиции фундаментальных решений, распределенных на носителе
функции $f(x)$, интенсивность которых определяется ее значением. Поэтому для построения решений,
удовлетворяющих определенным свойствам на бесконечности, следует вначале построить фундаментальные
решения, удовлетворяющие подобным свойствам. \vspace {3mm}

\textbf{4. Ударные волны как обобщенные решения уравнений. Условия на фронтах.}
Пусть $u(x')$- классическое решение в $R^N $, удовлетворяющее (17) - (18) на конечном числе
поверхностей типа $F$. Обозначим $\hat u(x,z)$ соответствующую регулярную обобщенную функцию: $\hat
u(x,z) =u(x')$, $x = (x_1' ,...,x_{N - 1}' ),\,z = x_N' $. Пользуясь (21), получим
    \[
\frac{{\partial \hat \sigma _{ij} }}{{\partial x'_j }} - \rho c^2 \frac{{\partial ^2 \hat u_i
}}{{\partial x_N ^2 }} + \rho G_i  = \left[ {\;\sigma _{ij} h_j  - \rho \,V^2 \;h_N \frac{{\partial
u_i }}{{\partial x_N }}} \right]_F \delta _F  + \quad
\]
\begin{equation}\label{(25)}
 + \frac{\partial }{{\partial x_j }}\left\{ {\left[ {\lambda \;u_k h_k \delta _{ij}
   + \mu (u_i h_j  + u_j h_i )} \right]_F \delta _F } \right\} -
   \frac{\partial }{{\partial x_N }}\left\{ {\left[ {\,u_i h_N \;} \right]_F \delta _F } \right\},
\end{equation}
где $a\delta _F $ - простой слой на $F$ , плотность которого определяется скачком перемещений на
фронте волны. В силу условий на фронтах ударных волн (16) и (18), правая часть (20) обращается в 0,
т.е. обобщенная функция $\hat u$ удовлетворяет тем же уравнениям (14), но уже в обобщенном смысле.

Отсюда,  следует простой формальный способ получения условий на скачки решений и их
производных на фронтах. Ударные волны являются обобщенными решениями уравнений (14), поэтому
достаточно приравнять плотности независимых слоев нулю, чтобы получить  условия на скачки на
фронтах ударных волн.\vspace{3mm}

\textbf{5. Фундаментальные решения. Матрица Грина}. При решении краевых задач важную роль играют
фундаментальные решения, которые описывают динамику среды при действии сосредоточенных источников
различного типа. Здесь приведем два фундаментальных решения, которые играют очень важную роль при
построении сингулярных граничных интегральных уравнений, разрешающих первую и вторую краевые задачи
теории упругости в случае бегущих нагрузок. В последующих статьях  мы приведем решение этих задач
во всем диапазоне скоростей движущихся нагрузок, за исключением звуковых, при которых стационарных
решений не существует.

Рассмотрим $\hat U_k^i $ - матрицу фундаментальных решений уравнения движения (14):
        \begin{equation}\label{(26)}
A_i^j \left( {\frac{\partial }{{\partial x'}}} \right)\hat U_j^k + \delta (x')\delta _i^k  =
0,\quad \quad i,j = \overline {1,N} ,
\end{equation}
удовлетворяющую условиям затухания на бесконечности:
    \begin{equation}\label{(27)}
\hat U_i^k  \to 0,\quad \partial_j\hat U_{i}^k  \to 0\,\, \textrm{при }\,\,x' \to \infty .
\end{equation}
Назовем  ее \emph{матрицей Грина} уравнения (14). Можно показать, что она является тензором.
При фиксированном $k$ ей соответствует
сосредоточенная сила, действующая  в направлении координатной  оси $X_k $, и бегущая со
скоростью $c$  вдоль оси $X_N $.

 Для произвольной регулярной  силы $G$ соответствующее решение имеет вид свертки
        \begin{equation}\label{(27)}
\hat u_i  = \hat U_i^k *\hat g_k  = \int\limits_{R^N } {U_i^k (x' - y)} g_k (y)dy_1...dy_N.
\end{equation}

Приведем  вид $\hat U_i^k$ для разных скоростей $c$, который получен  с использованием преобразования
Фурье,  имеющего  вид [7]:
\begin{equation}\label{(28)}
\bar U_i^j  = \frac{{M_2^2\delta _{ij} }}{{(\left\| \xi \right\|^2  - M_2^2 \xi _N^2 )}} -
\frac{{\xi _i \xi _j }}{{ \xi _N^2 }}\left( {\frac{1}{{\left\| \xi  \right\|^2  - M_2^2 \xi _N^2 }}
- \frac{1}{{\left\| \xi  \right\|^2  - M_1^2 \xi _N^2 }}}, \right)
\end{equation}
Откуда следует:\[
U_i^j (x,z) = M_2^2 \delta _i^j f_{02} (\left\| x \right\|,z) +\left( {f_{21} ,_{ij} (\left\| x
\right\|,z) - f_{22} ,_{ij} (\left\| x \right\|,z)} \right),
\]
где $f_{km} $ - оригиналы функции  $\hat f_{km}  = \frac{{\xi _N^{ - k} }}{{\left\| \xi  \right\|^2
- M_m^2 \xi _N^2 )}}$, вид которых существенно зависит от размерности задачи и чисел Маха (см.[2]).

Так $\bar f_{0m}  = \left( {\left\| \xi  \right\|^2  - M_m^2 \xi _N^2 } \right)^{ - 1}.$
Легко видеть, что   это преобразование Фурье фундаментального решения уравнения
$$\Delta _{N - 1} \hat f_{0m}  + (1 - M_m^2 )\hat f_{0m}  + \delta (x)\delta (z) = 0.$$
При дозвуковых скоростях это уравнение эквивалентно эллиптическому
уравнению Лапласа-Пуассона, при сверхзвуковых - волновому
уравнению (уравнению Даламбера для N=2), при звуковой скорости
переменная z  в уравнении исчезает, уравнение становится
параболическим, т.к. размерность пространства на единицу выше. Что
и определяет тип уравнений (15), отмеченный выше.

При построении
фундаментальных решений,  помимо условий затухания
решений на бесконечности, учитывались условия излучения  [4,7]:\\
при $ {M_m < 1} $:
$f_{km}  = \theta (z)\int\limits_0^z {f_{(k - 1)m} (x,z)dz}  - \theta ( - z)\int\limits_z^0 {f_{(k -
1)m} (x,z)dz}$ ,\\
при $ {M_m  > 1} $: $ \quad f_{km}  = \theta (z)\int\limits_0^z {f_{(k -
1)m} (x,z)dz}, \quad k = 1,2. $

\emph{Пространственная задача ( $N=3$)}. При дозвуковых скоростях ($M_k<1$):
\[
\;4\pi f_{ok} (\left\| x \right\|,z) = (m_k^2 \left\| x \right\|^2 + z^2 )^{ - 1/2} ,\;\;4\pi
f_{1k}  = {\mathop{\rm sgn}} \;\,\left| {z\;} \right|\;ln\left( {(\,\left| z \right| + V_k^ +
)/m_k \left\| x \right\|} \right),
\]
\[
4\pi f_{2k}  = \left| z \right|\;ln\left( {(\left| z \right| + V_k^ +  )/m_k \left\| x \right\|}
\right) - V_k^ +   + m_k \left\| x \right\|;
\]
при звуковых скоростях($M_k=1$):
\[
2\pi \;f_{ok} (\left\| x \right\|,z) =  - \delta (z)\ln \left\| x \right\|,\quad 2\pi \;f_{1k}  =
\, - \theta (z)\;ln\left\| x \right\|, \,\,\,2\pi \;f_{2k}  =  - \,z\,\theta (z)\;ln\left\| x
\right\|,
\]
при сверхзвуковых скоростях($M_k>1$):
\[
2\pi \;f_{ok} (\left\| x \right\|,z) = \theta (z - m_k \left\| x \right\|)(z^2  - m_k^2 \left\| x
\right\|^2 )^{-1/2} ,
\]
\[
\;2\pi f_{1k}  = \theta (z - m_k \left\| x \right\|)\;\ln \;\left( {\left( {z + V_k^ -  }
\right)/m_k \left\| x \right\|} \right),
\]
\[
2\pi f_{2k}  = \theta (z - m_k \left\| x \right\|)\left( {z\ln \left( {\left( {z + V_k^ -  }
\right)/m_k \left\| x \right\|} \right) - V_k^ -  } \right).
\]
Здесь введены обозначения: $\theta \left( z \right)$ -- функция Хевисайда,
$$ V_k^ +   = (z^2
+ m_k^2 \left\| x \right\|^2 )^{1/2} ,\quad V_k^ -   = (z^2  - m_k^2 \left\| x \right\|^2 )^{1/2}
,\quad \theta _k  = \theta \left( {z - m_k \left\| x \right\|} \right).$$

 \emph{Плоская задача ($N=2$)}. При дозвуковых скоростях ($M_k<1$):
\[
2\pi m_k f_{ok} (\left| x \right|,z) =  - ln V_k^{+},
\quad
2\pi m_k f_{1k}  =   -|z|\ln V_k^{+} + |z| - m_k |x| arctg \frac{|z|}{ {m_k |x|} },
\]
\[
  4\pi m_k f_{2k}  =   -(V_k^{+})^2\ln V_k^{+} + 1,5z^2 +
2m_k |x|(m_k |x|-|z|)\,arctg  \left|  \frac{z}{{m_k x} }\right|   ;
\]
при звуковых скоростях ($M_k=1$):
    \[
2f_{ok} (\left| x \right|,z) =  - \delta (z)\left| x \right|,\quad 2f_{1k}  = \,\theta (z)\left| x
\right|,\quad 2f_{2k}  = z\,\theta (z)\left| x \right|;
\]
при сверхзвуковых скоростях  ($M_k>1$):
\[
2m_k f_{ok} (\left| x \right|,z) = \,\theta (z - m_k \left| x \right|), \quad 2m_k f_{1k}  = (z -
m_k \left| x \right|)\theta (z - m_k \left| x \right|),
\]
\[
4m_k f_{2k}  = (z - m_k \left| x \right|)^2 \theta (z - m_k \left| x \right|).
\]

    Построенные тензора обладают следующими свойствами симметрии:
            \begin{equation}\label{(29)}
\hat U_j^i (x,z) = \hat U_i^j (x,z) = \hat U_j^i ( - x,z).
\end{equation}
  Легко выделяются объемные и сдвиговые составляющие тензора  $\hat U$:
          \[
\hat U_j^i  = \hat U_{j1}^i  + \hat U_{j2}^i ,
\]
        \begin{equation}\label{(29)}
\hat U_{j1}^i  = c^{ - 2} f_{21'ij}^{} ,\quad \hat U_{j2}^i  = c_2^{ - 2} \delta _j^i f_{02}  - \nu
^{ - 2} f_{22'ij}.
\end{equation}
$\hat U_{j1}^i $ описывает деформации сжатия-растяжения, которые распространяются в упругой среде
со скоростью  $c_1 $,  а $\hat U_{j2}^i $-- сдвиговые деформации, которые распространяются
медленнее со скоростью $c_2 $.

 В сверхзвуковом случае носителем функций является конус
$z>m_k\|x\|$.  Последнее является \textit{условием излучения}, т.к. из физических
соображений вне этого конуса перемещения  упругой среды отсутствуют в силу конечности
скорости распространения возмущений, которая не может быть выше  соответствующей
звуковой для определенного типа деформаций. Заметим, что эти условия появляются
 в результате построения решения задачи для объемной и сдвиговой составляющей тензора Грина
 отдельно в зависимости от скорости бегущей нагрузки.

На  поверхности $z=m_k\|x\|$ в трехмерном случае тензор Грина имеет
бесконечную особенность, в плоском случае - конечный скачок [7].

\vspace{3mm}

\textbf{6. Фундаментальная матрица напряжений $\hat {\rm T}_i^j $}. Используя закон Гука (1), введем
тензоры напряжений, порождаемые тензором Грина:
 \begin{equation}\label{(31)}
\begin{array}{l}
 \hat S_{ij}^k (x,z) = \lambda \delta _{ij}\partial_m\hat U_m^k
  + \mu \left( {\partial_j \hat U_i^k   + \partial_i\hat U_j^k  } \right), \\
 \hat \Gamma _i^k (x,z,n) = \hat S_{ij}^k (x,z)n_j ,\quad
 \hat {\rm T}_i^j (x,z,n) =  - \hat \Gamma _j^i (x,z,n).\quad  \\
 \end{array}
\end{equation}
Верна следующая теорема.

Т е о р е м а 2. \emph{Тензор  $\hat {\rm T}_i^j $  является  обобщенным решением  уравнения}:
\[
A_i^j \left( {\partial _{x'} } \right)\hat T_j^k  =  - K_k^i \left( {\partial _{x'} ,n}
\right)\delta (x').
\]
\emph{где }   $K_i^l \left( {\partial _{x'} ,n} \right) = \lambda n_i
\partial _l  + \mu _j \left( {\delta _i^l \partial _j  + \delta
_j^l \partial _i } \right)$.

    Д о к а з а т е л ь с т в о: Из (32), (1) получим:
$ - \hat {\rm T}_i^j  = \hat \Gamma _i^j  = K_i^l \left( {\partial _{x'} ,n} \right)\hat U_l^j (x,z)
$.
 Следовательно,
$
A_i^j \left( {\partial _{x'} } \right)\hat T_j^k  =  - A_i^j \left( {\partial _{x'} } \right)\hat
\Gamma _k^j  =  - A_i^j \left( {\partial _{x'} } \right)K_k^l \left( {\partial _{x'} ,n}
\right)\hat U_l^j  =
 =  - K_k^l \left( {\partial _{x'} ,n} \right)A_i^j \left( {\partial _{x'} } \right)
 \hat U_j^l (x,z) =  - K_k^l \left( {\partial _{x'} ,n} \right)\delta _i^l \delta (x') =
  - K_k^i \left( {\partial _{x'} ,n} \right)\delta (x').$
Теорема доказана.

 Тензор вычисляется, используя соответствующие формулы п.4 в
зависимости от размерности задачи и скорости движения. Аналогично можно выделить объемную и
сдвиговую составляющие $\hat {\rm T}_i^j (x,n) = \hat {\rm T}_{i1}^j  + \hat {\rm T}_{i2}^j
$. Он описывает динамику упругой среды при действии бегущей сосредоточенной  силы
мультипольного типа. \vspace{3mm}

\textbf{Заключение.} Представленные фундаментальные матрицы можно использовать для изучения
динамики сред с бегущими нагрузками, распределенными по объему, поверхности или прямой, используя
формулу теоремы 1.

Если  $g_k$ - локально интегрируемые функции с конечным носителем, что типично для физических
задач, в силу регулярности $U$,
 решение  имеет вид:
\[
\hat u_i (x,z) = \int\limits_{{\rm supp}G(y)} {U_i^k (x' - y)}g_k (y)dy_1 ...dy_N.
\]
Условия на $g_k$  можно ослабить, если учесть асимптотику $U$ на бесконечности.

Если бегущая сила сосредоточена на цилиндрической поверхности $S$, образующая которой параллельна
оси $Z$, то ее можно описать сингулярной функцией с компонентами $\hat g_k  = g_k (x,z)\delta _S
(x,z)$. Тогда решение имеет вид:
$$\hat u_i (x,z) = \int\limits_S {U_i^k (x' - y)} g_k (y)dS(y).$$

Если бегущая сила сосредоточена на оси $Z$, ее можно описать сингулярной функцией $\hat g_k  = g_k
(z)\delta (x)$. Тогда решение
 имеет вид:
\[
\hat u_i (x,z) = \int\limits_{R^1 } {U_i^k (x,z - y)} g_k (z)dz.
\]
Используя вид $U$  в зависимости от размерности задачи и скорости бегущей нагрузки легко получить
аналитические формулы для определения перемещений среды.
При этом эти решения, в силу леммы Дюбуа-Реймона [1],  будут классическими.

В случае сингулярных сил другого вида для построения обобщенного решения
 следует использовать формулу свертки   согласно ее определению для обобщенных функций (см.[1,2]).

\vspace{3mm}

\emph{Ключевые слова}: упругая среда, уравнения Ламе, бегущие нагрузки, фундаментальные решения,
сверзвуковая, трансзвуковая, дозвуковая скорость, ударные волны. \vspace{3mm}

\centerline{\textbf{Список использованных источников}}

1. Владимиров В.С. Уравнения математической физики. М. 1978. 528с.

2.Хермандер Л. Анализ линейных дифференциальных операторов с частными производными.
Т.1. Теория распределений и анализ Фурье. - М.1986.462 с.

3. Алексеева Л.А. Обобщенные решения краевых задач для одного класса бегущих решений
волнового уравнения // Математический журнал. Т.8, №2(28). 2008.  C.1-19.

4. Алексеева Л.А.Фундаментальные решения в упругом пространстве в случае бегущих нагрузок //
 Прикладная математика и механика. 1991. Т.55. №5. С.854-862.

 5. Новацкий В. Теория упругости. М., 1975. 872 с.

6. Петрашень Г.И. Основы математической теории распространения упругих волн.-
В кн. Вопросы динамической теории распространения сейсмических волн. Л. 1978. Вып.XVIII, 248с.

7. Алексеева Л.А. Граничные интегральные уравнения краевых задач для класса стационарных бегущих
 решений волновых уравнений в цилиндрических областях/ Препринт Института математики МН-АН РК.1997.72с.

 \vspace{3mm}

\newpage

 Алексеева Л.А. Обобщенные решения уравнений Ламе в случае
бегущих нагрузок. Ударные волны//Математический журнал. Т.9.
2009.№1 (31).С.16-25.
\vspace{3mm}

Исследуется система уравнений Ламе, описывающая
движение упругой среды при дозвуковых, транс- и
сверхзвуковых скоростях движения источника возмущений, и ее
решения в пространстве обобщенных вектор-функций. Рассмотрены
вопросы, связанные с возникновением ударных волн, которые
возникают в среде при сверхзвуковых источниках возмущений. На
основе теории обобщенных функции предложен метод определения
условий на скачки решений и их производных на фронтах ударных
волн.
\vspace{3mm}

Alexeyeva L.A. The generalized solutions  of the Lama's
equations  in the case of  running loads. The shock
waves//Mathematical journal. V.9. 2009. No 1 (31).P.16-25.
\vspace{3mm}

 The system of
Lama's equations is investigated, describing the motion of the
elastic media under subsonic, transonic and supersonic velocities
of the moving source of distributions, and its decisions in space
of generalized vector-functions. The questions are considered
connected with arising shock waves, which appear in ambience under
supersonic source of distributions. On base of the generalized
functions theories the method of the determination of the
conditions on gaps of the decisions and their derivatives on shock
waves fronts is offered. \vspace{3mm}

\vspace{3mm}

\end{document}